# Prediction of the Subjective Impression of Passenger Car Roll Dynamics on the Driver Based on Frequency-Domain Characteristic Values

Dipl.-Ing. Andreas Apfelbeck, Dipl.-Ing. Stefan Wegner, apl. Prof. Dr.-Ing. Roman Henze, Prof. Dr.-Ing. Ferit Küçükay

*BMW Group, München, Germany*

Dipl.-Ing. Andreas Apfelbeck, BMW Group, Knorrstraße 147, 80788 München, Germany, andreas.apfelbeck@bmw.de

Dipl.-Ing. Stefan Wegner, BMW Group, Knorrstraße 147, 80788 München, Germany

apl. Prof. Dr.-Ing. Roman Henze, Institut für Fahrzeugtechnik, TU Braunschweig, 38106 Braunschweig, Germany

Prof. Dr.-Ing. Ferit Küçükay, Institut für Fahrzeugtechnik, TU Braunschweig, 38106 Braunschweig, Germany

# Prediction of the Subjective Impression of Passenger Car Roll Dynamics on the Driver Based on Frequency-Domain Characteristic Values

Characteristic values are essential for the design and assessment of driving dynamics during the early stages of the development process of passenger cars. Compared to other aspects of vehicle dynamics however, the relationship between measurable parameters and the subjective perception of vehicle roll dynamics has not been researched extensively. In this paper, a study is presented in which several variants of a vehicle with an electronically controlled suspension were rated by test subjects regarding its roll dynamics and measured in a standardized driving manoeuvre. The resulting subjective ratings and objective characteristic values are then used to derive models to predict the subjective liking of several roll dynamics aspects based on objective frequency-domain parameters. Finally, the resulting prediction models are validated using measurements of additional vehicles.

Keywords: roll dynamics; objective evaluation; subjective evaluation; prediction

**Introduction**

In the early stages of the development process of a new vehicle, simulation often is the only development tool available to engineers because actual prototype vehicles do not exist yet. And, with the increasing performance and accuracy of simulation methods and models as well as the decreasing amount of physical prototype vehicles built during the development of a new car, simulations will be used even more in the future.

In this simulation driven phase, prediction models are essential for the assessment and goal-oriented design of the driving dynamics of a vehicle in order to make the driving characteristics of the virtual prototype converge as much as possible towards the desired driving characteristics of the final series production vehicle, e.g. [1]. Additionally, those predictors help in assessing the current development stage as well as changes between iterative development steps. The predictors can be derived based on experience, by using predecessor vehicles or by employing a methodical approach. The latter is when methods of the objective evaluation of vehicle dynamics are used.

By now, there have been several attempts to describe the individual aspects of vehicle dynamics using characteristic values derived from standardized driving manoeuvres. The focus of those studies was often laid on lateral dynamics or steering feel (e.g., [2 – 4]) as these aspects are intuitively relevant for a driver's subjective impression of vehicle dynamics.

Despite the large influence of the roll dynamics of a vehicle on the subjective driving impression of its passengers, to this day it was only addressed in a small number of publications (cf. [4 – 10]) with merely a few studies explicitly picking the roll dynamics as a more central theme (cf. [11 – 17]). However, in most of these publications only some partial aspects were broached without focusing too much on how the driver's perception of roll dynamics can actually be predicted. In short, there are still no reliable models for the assessment of a vehicle's roll dynamics based on objective CVs as of yet, which is why at this point in time no reliant objective assessment of roll dynamics is possible, which motivated the research presented in the following.

This paper starts with the description of a study which was conducted to obtain subjective ratings of several different roll dynamics variants of a BMW X5. The same variants were also measured in a standardized driving manoeuvre to get objective measurement data. After pre-processing both subjective ratings and objective measurement data and selecting characteristic values from the latter, correlation analyses were performed to validate the data sets and to obtain initial findings. Subsequently, regression analyses were used to derive more complex models for the prediction of subjective ratings of roll dynamics aspects. Finally, the models were validated using measurements of additional vehicles.

**Objective Evaluation and Subjective Assessment of Vehicle Roll Dynamics**

The general approach regarding the objective evaluation of vehicle dynamics is fairly well established by now. Basically, several vehicle variants which differ in the research aspects are measured using standardized driving manoeuvres and rated by several test subjects. The resulting objective and subjective data is analysed using descriptive and analytical statistics in order to establish reliable relations between the subjective ratings and the CVs extracted from the objective measurement data. Furthermore, regression analyses can be used to derive more detailed but also more complex models for the prediction of the liking of certain vehicle aspects.

*Study Design*

To obtain the subjective ratings of the roll dynamics variants, a single-blind study with 16 test subjects was conducted on the BMW proving grounds in Aschheim near Munich. The tests subjects were asked to rate several variants of a BMW X5 with electronically controlled dampers, electronically controlled anti-roll bars and electric power steering. The roll dynamics of the car was modified by changing the calibration of the above-mentioned control systems. By using only one vehicle in the study instead of several different ones, the risk of the test subjects subconsciously including additional effects while rating the roll dynamics could be reduced.

Overall, the 16 test subjects assessed seven criteria of six vehicle variants which in total amounted to 1824 individual ratings. Specifically, the following vehicle variants were used, each variation described in relation to the reference variant:

- A reference variant similar to the production vehicle version of the BMW X5 (variant RV).
- Two variants with increased and decreased roll damping (variants RD↑ and RD↓, respectively).
- Two variants with decreased and increased level of steering support (variants ST↑ and ST↓, respectively).
- A variant with decreased eigenfrequency of the transfer function from lateral acceleration to roll angle (EF↓) which essentially increased the roll angle at lower excitation frequencies.

All test subjects could be considered expert drivers who were working in the development department of driving dynamics at BMW Group in Munich and had extensive experience in rating vehicle dynamics. Each test subject started with the reference variant followed by an individually rearranged order of the remaining variants to avoid familiarization effects over the course of the assessments, the only exception being the reference variant which was placed at the beginning and the end of each subject's assessment run in order to be able to check their repetition accuracy.

The test subjects were asked to evaluate the following three aspects for each criterion and variant:

- Perceived intensity of the criterion.
- Subjective liking.
- Direction of improvement.

The intensity rating (also called quasi-objective rating) is used to verify the assessment quality of the test subjects, i.e., if the perceived subjective change of the vehicle variant is consistent with the actual change in vehicle dynamics, and also to select more robust

CVs from the correlation analysis.

Figure 1: Sample page from the questionnaire used in the study.

The questionnaire used in the study is shown in figure 1. The "BI-Skala" was chosen because it is fairly common in the automotive industry, which is why the test subjects were already familiar with it. The scale ranges from 1 to 10 but was modified to use a finer resolution in the interval which most of the ratings were expected to end up in. Additionally, by using an absolute scale in contrast to a relative one, the results of the study are more universally applicable because they can be used for the estimation of actual subjective ratings instead of only a relative order of variants.

The following seven criteria were rated by the test subjects:

- Absolute roll angle at lower excitation frequencies (RAL).
- Absolute roll angle at higher excitation frequencies (RAH).
- Time delay between driver input and vehicle roll motion at lower excitation frequencies (TDL).

- Time delay between driver input and vehicle roll motion at higher excitation frequencies (TDH).

- Initial roll motion of the vehicle body at the beginning of the driver input (IRM).

- Overshoot of the body roll motion at the end of the driver input (ROS).

- Overall rating for the roll dynamics of the variant (OR).

A two-lane straight was selected as test track which the subjects could perform their driving manoeuvres on. The standardized manoeuvres were recorded on the same track. The test subjects were instructed to evaluate the vehicle from a customer's perspective (i.e., up to a maximum lateral acceleration of 4 m/s²) at a velocity of about 100 km/h because the objective measurements were performed at the same vehicle speed. The study conductor was not present in the car during the evaluation of each variant so as not to influence the assessor's driving behaviour. There was no predetermined time span during which the subjects had to assess a variant. Instead, they could decide on their own how much time they needed for their assessment. After each evaluation round they left the test track to fill out the questionnaire while the study conductor set up the next variant by changing the calibration of the car's control systems. This procedure was repeated until all variants had been rated.

*Statistical Analysis*

Prior to the actual analysis of the subjective data, the ratings of each test subject $x_i$ were standardized using the mean value $\mu$ and the standard deviation $\sigma$ of the test subjects' ratings:

$$z_i = \frac{x_i - \mu}{\sigma} \quad (1)$$

The standardized data set was then inversely transformed onto the mean value $\mu_{tot}$ and the standard deviation $\sigma_{tot}$ of the data set of all 16 test subjects, mapping the ratings measured on each test subject's individual scale onto a common rating scale:

$$x_i = \sigma_{tot} \cdot z_i + \mu_{tot} \qquad (2)$$

Outliers were detected using Tukey's fences where a value is treated as an outlier if its distance to the quartiles Q1 and Q3 exceeds 1.5 times the interquartile range (IQR) where Q1 is 25th percentile and Q3 is the 75th percentile. Instead of just trimming the outlier ratings from the data set, they were winsorized, limiting each value to the Tukey fence value it was closest to in order to not completely disregard the information contained within those values.

Correlations were assumed to be statistically significant if the corresponding p-values were less than the significance level $\alpha = 0.05$. In this case, correlation coefficients listed in any subsequent table are printed in boldface. Finally, regression analyses were performed to identify equations which describe statistically significant (nontrivial) relationships between the measured objective data and the subjective ratings. During the stepwise linear regressions, terms were added and removed iteratively to the model equations in order to increase the value of the adjusted R-squared $R^2_{adj}$.

*Objective Measurement Data and Characteristic Values*

To obtain reliant objective data, an inertial measurement unit and a dedicated measurement steering wheel were used to record the vehicle data during the standardized driving manoeuvres. The manoeuvre used in this analysis is the continuous sine steering (CSST) which provides information about the frequency-domain behaviour of a vehicle. For the CSST manoeuvre the steering frequency is continuously increased while vehicle speed and steering wheel angle amplitude are kept constant, starting with a predetermined amplitude for a given stationary value of the lateral acceleration. In this case a vehicle speed of $v = 100 \: km/h$ and a lateral acceleration of $a_y = 4 \: m/s^2$ were used.

The transfer function plots can then be estimated from the measurement data using Welch's averaged modified periodogram which yields the quotient of the cross power spectral density (CPSD) of the signals used. The magnitude of the CPSD represents the Bode magnitude plot and the phase angles of the CPSD represent the Bode phase plot of the transfer function $G(s) = Y(s)/U(s) = G_{uy}(s)$, where $U$ is the system input and $Y$ is the system output.

A simplified chain of effects for the roll dynamics of a vehicle is shown in equation (3)

$$M_H \to \delta_H \to (\ddot{\psi}, \dot{\beta}) \xrightarrow{\int dt} (\dot{\psi}, \beta) \to a_y \to \ddot{\varphi} \xrightarrow{\int dt} \dot{\varphi} \xrightarrow{\int dt} \varphi \qquad (3)$$

where $M_H$ is the driver steering torque, $\delta_H$ is the steering wheel angle, $\psi$ is the yaw angle, $\beta$ is the slip angle, $a_y$ is the lateral acceleration and $\varphi$ is the vehicle roll angle. In contrast to previous research, the steering torque was also included, so that no potentially influencing variable between driver input and vehicle output is neglected.

Based on equation 3, the transfer functions from the excitation states $M_H$, $\delta_H$ and $a_y$ to the roll angle $\varphi$ and its time derivatives were selected and the values of the amplitude gains $V$ and phase angles $\phi$ at the frequencies 0.6, 0.9, 1.2 and 1.5 Hz determined. Additionally, the following parameters were used in the analysis:

- Quasi-stationary value of the amplitude gain at a frequency $f = 0.3 \: Hz$:

$$V_{uy}^0 = |G_{uy}(0.3)|$$

- Maximum amplitude gain of the frequency response:

$$V_{uy}^{max} = \max|G_{uy}(j\omega)|$$

- Magnification factor $\beta$:

$$\beta_{uy} = \frac{\max|G_{uy}(j\omega)|}{V_{uy}^0}$$

- Position of the maximum magnification of the frequency response of $G_{uy}(j\omega)$:

$$\omega_{uy}^0$$

Because the phase angles of the frequency response are shifted by $+90°$ for each differentiation of the output signal with respect to time, no further information is contained in additional phase response plots. Thus, only the phase plot of the transfer function $G_{uy}(s)$ is shown in figure 2 and only the phase angle values from the transfer functions to the roll angle $\varphi$ are used.

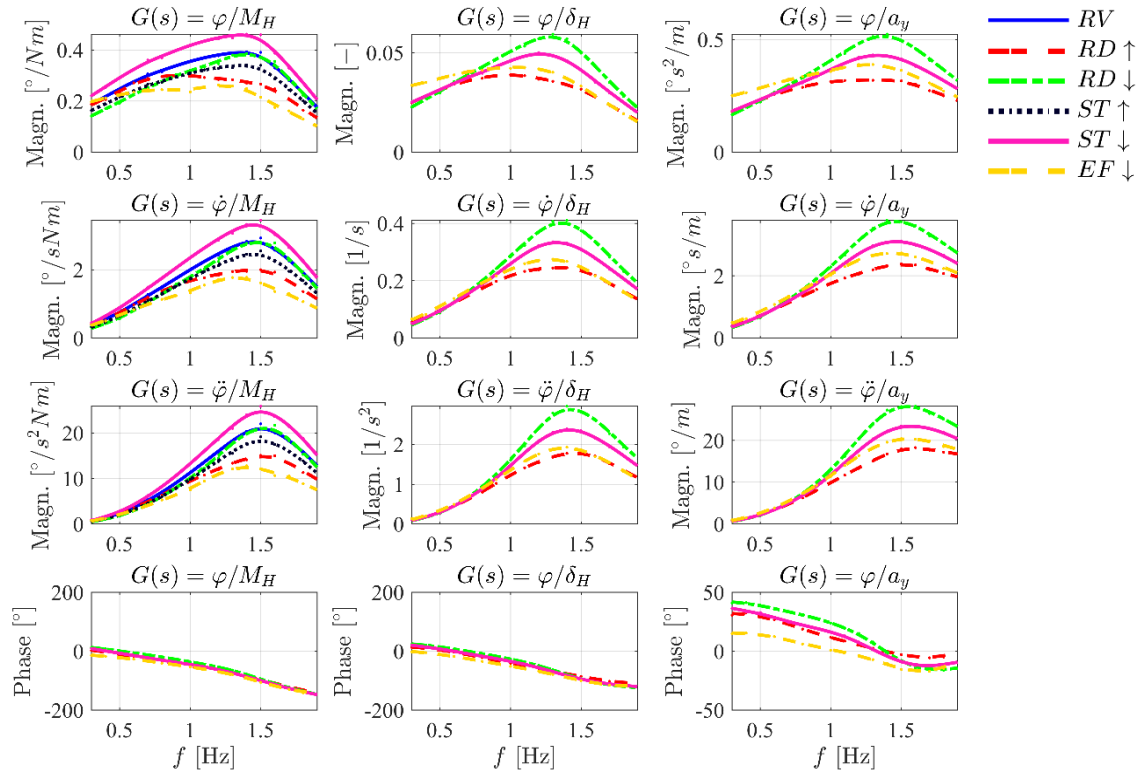

Figure 2: Frequency response of the BMW X5 variants.

**Results**

In the following, the data obtained through the before-mentioned methods is presented. The averaged ratings of the variants are discussed and the effects of the roll dynamics variations on the subjective ratings are analysed. The measurement data is used to determine the frequency response of the roll dynamics system represented by the Bode plots shown in figure 3. Characteristic values are then selected and along with the subjective ratings analysed for correlations. Finally, regression models are derived to predict the subjective ratings based on these CVs.

*Subjective Ratings*

The individual subjective ratings of each test subject were standardized and then transformed onto the mean and the variance of the ratings of the whole group of test subjects as described in the previous section. The resulting averaged standardized ratings of each variant and criterion are shown in figure 3 and table 1 (with the background colours denoting the changes relative to the reference variant, where a green background colour equals higher ratings and a red background colour equals lower ratings compared to RV) and described in the following.

Table 1: Averaged standardized liking ratings of each variant.

| | | Criterion | | | | | | |
|---|---|---|---|---|---|---|---|---|
| | | RAL | RAH | TDL | TDH | IRM | ROS | OR |
| Variant | RV | 7.8 | 6.9 | 7.4 | 7.0 | 7.6 | 7.7 | 7.3 |
| | RD↑ | 7.6 | 7.5 | 7.1 | 7.3 | 7.3 | 7.4 | 7,1 |
| | RD↓ | 7.6 | 6.4 | 7.4 | 6.7 | 7.1 | 7.2 | 6,7 |
| | ST↑ | 7.8 | 7.1 | 7.7 | 7.2 | 7.7 | 7.6 | 7,4 |
| | ST↓ | 7.7 | 6.9 | 7.4 | 6.9 | 7.5 | 7.4 | 7,1 |
| | EF↓ | 7.3 | 7.2 | 7.0 | 7.1 | 7.0 | 7.4 | 6,9 |
| $\Delta_{min/max}$ | | 0.5 | 1.1 | 0.7 | 0.6 | 0.7 | 0.5 | 0.7 |

RAL = roll angle at lower frequencies; RAL = roll angle at higher frequencies;
TDL = time delay at lower frequencies; TDH = time delay at higher frequencies;
IRM = initial roll motion; ROS = roll overshoot; OR = overall rating

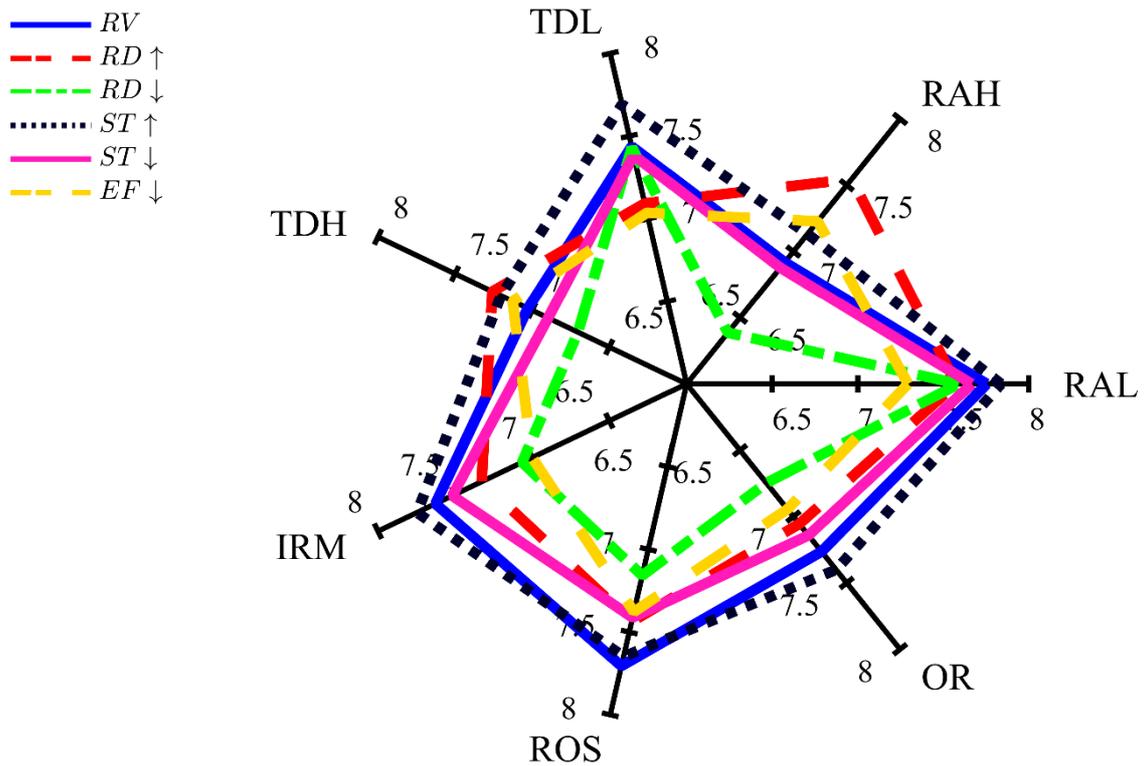

Figure 3: Spider plot of the subjective liking ratings.

The increase of roll damping (variant RD↑) resulted in a rating improvement of both RAH and TDH but decreased the ratings of RAL, TDL, IRM as well as OR. The reduction of roll damping (variant RD↓) on the other hand lowered the ratings for all criteria except for TDL which remained constant. The reduction of steering support (variant ST↑) improved the ratings of all criteria, especially TDL, except for ROS. In contrast, the increase of steering support (variant ST↓) reduced the ratings of all criteria. The modification of the roll eigenfrequency (variant EF↓) resulted in reduced ratings of every criterion but RAH and TDH.

The variant with increased steering torque (ST↑) turned out to receive the highest ratings overall with six out of the seven criteria rated higher than the reference variant. This emphasizes the importance of steering feel for the perception of roll dynamics. The remaining variants were rated worse compared to the reference variant, with variant RD↓ being the worst, followed by variant EF↓. The biggest rating improvement of an individual criterion could be observed for RAH of the variant RD↑ (by 0.6 BI), whereas the biggest rating reduction could be observed for IRM of the variant EF↓ (by -0.6 BI).

Table 2: Subjective intensity ratings of each criterion.

|  |  | Criterion |  |  |  |  |  |
|---|---|---|---|---|---|---|---|
|  |  | RAL | RAH | TDL | TDH | IRM | ROS |
| Variant | RV | 2.4 | 3.2 | 2.5 | 3.1 | 2.4 | 2.5 |
|  | RD↑ | 2.7 | 2.6 | 2.7 | 3.1 | 2.4 | 2.7 |
|  | RD↓ | 2.5 | 3.9 | 2.6 | 3.5 | 3 | 3 |
|  | ST↑ | 1.7 | 3.1 | 2.2 | 3 | 2.2 | 2.5 |
|  | ST↓ | 2.4 | 3.6 | 2.5 | 2.9 | 2.6 | 2.8 |
|  | EF↓ | 2.7 | 3.4 | 2.7 | 2.8 | 2.9 | 2.7 |
| $\Delta_{min/max}$ |  | 1.0 | 1.3 | 0.5 | 0.7 | 0.8 | 0.5 |

The intensity ratings of the six variants are shown in figure 4 and in table 2 (where higher ratings equal higher background colour intensities). They are, for the most part, inversely proportional to the liking ratings except for the criterion TDH (cf. subsequent correlation analysis in section "Results" for detailed information about the correlation coefficients).

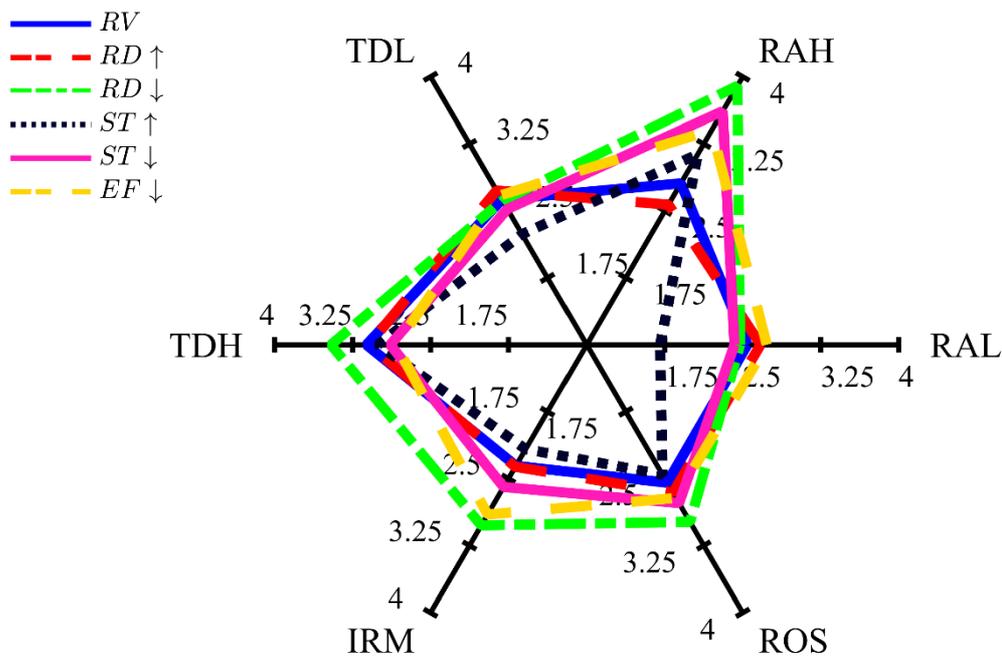

Figure 4: Subjective intensity ratings of the vehicle variants.

Also, compared to the other criteria, the rating differences of both the time delay criteria TDL and TDH and of the criterion ROS between the variants are small. The effect of a higher steering torque (less excitation of the vehicle lateral and thus roll dynamics by the driver) can be clearly observed from the intensity ratings of RAL, TDL and IRM, which were perceived to be smaller than the reference variant's.

Based on the dataset of the liking ratings shown in table 1, a two-pooled two-sample t-test for unequal means and unequal variances at a significance level $\alpha = 0.05$ was used to examine which vehicle variants could be distinguished with statistical significance by the test subjects. As could be expected from the artificial manipulation of individual vehicle dynamics properties, the test indicated that only a few criteria of the different variants could be distinguished with statistical significance regarding the subjective ratings, cf. table 3. The results obtained for TDL, TDH, IRM, ROS, and OR thus need to be interpreted with care as further analysis might be required. As to the criteria RAL and RAH, the test subjects distinguished the combinations shown in table 3 with statistical significance.

Table 3: Liking ratings distinguished with statistical significance.

|  |  | Variant | | | | | |
|---|---|---|---|---|---|---|---|
|  |  | **RV** | **RD↑** | **RD↓** | **ST↑** | **ST↓** | **EF↓** |
| **Variant** | **RV** |  |  |  |  |  |  |
|  | **RD↑** | - |  |  |  |  |  |
|  | **RD↓** | - | RAH |  |  |  |  |
|  | **ST↑** | - | - | RAH |  |  |  |
|  | **ST↓** | - | RAH | - | - |  |  |
|  | **EF↓** | - | - | RAH | RAL | - |  |

Regarding the actual rating changes, the modification of roll damping primarily affected the rating of RAH. This is consistent with expectation because the variation of roll damping has its greatest effect on the roll gain at the roll eigenfrequency of a vehicle which is usually located at higher excitation frequencies. Typically, a driver excites the vehicle body in the lower frequency area well below 0.6 Hz, which significantly reduces the effect of any variation of roll damping.

The modification of steering torque (variants ST↑ and ST↓) affected all rating criteria approximately equally, a reduction of steering support leading to better ratings of the criteria and vice versa. Because the driver needs less force to excite the vehicle with increasing steering support, he in this case also tends to use greater steering wheel angles while controlling the vehicle compared to a vehicle with less steering support. This in turn leads to more roll motion which is interpreted as a less supported vehicle body and ultimately reflected in the ratings.

The eigenfrequency modification (variant EF↓) had a primary effect on RAL, TDL, IRM, ROS and OR, i.e., on every criterion but the higher-frequency ones. Compared to the other variants, the ratings of this variant deviated the most from the reference variant which again shows the importance of lower excitation frequencies for the perception of roll dynamics.

No statistically significant differences were found between the ratings of the reference variant and its repetition, which confirms the assessment qualification of the test subjects.

*Objective Parameters*

The numerical values of the previously presented CVs are listed in table 4. The individual amplitude gain and phase angle values were omitted for the sake of clarity. Overall, 126 CVs have been selected for each vehicle variant from the measurement data.

Table 4: Selected frequency-domain CVs of the six vehicle variants.

| Transfer function $G_{uy}$ | CV | Variant | | | | | | Unit |
|---|---|---|---|---|---|---|---|---|
| | | RV | RD↑ | RD↓ | ST↑ | ST↓ | EF↓ | |
| $G_{M_H\varphi}$ | $V_{uy}^0$ | 0.18 | 0.18 | 0.17 | 0.18 | 0.18 | 0.25 | °/Nm |
| | $V_{uy}^{max}$ | 0.43 | 0.32 | 0.52 | 0.43 | 0.43 | 0.39 | °/Nm |
| | $\omega_0^{uy}$ | 1.33 | 1.32 | 1.37 | 1.33 | 1.33 | 1.26 | Hz |
| | $\beta_{uy}$ | 2.36 | 1.77 | 3.10 | 2.36 | 2.36 | 1.59 | − |
| $G_{M_H\dot\varphi}$ | $V_{uy}^0$ | 0.36 | 0.38 | 0.34 | 0.36 | 0.36 | 0.47 | °/(sNm) |
| | $V_{uy}^{max}$ | 3.09 | 2.35 | 3.73 | 3.09 | 3.09 | 2.72 | °/(sNm) |
| | $\omega_0^{uy}$ | 1.46 | 1.51 | 1.46 | 1.46 | 1.46 | 1.43 | Hz |
| | $\beta_{uy}$ | 8.63 | 6.21 | 11.06 | 8.63 | 8.63 | 5.75 | − |
| $G_{M_H\ddot\varphi}$ | $V_{uy}^0$ | 0.64 | 0.73 | 0.62 | 0.64 | 0.64 | 0.82 | °/($s^2$Nm) |
| | $V_{uy}^{max}$ | 23.3 | 18.0 | 28.0 | 23.3 | 23.3 | 20.3 | °/($s^2$Nm) |
| | $\omega_0^{uy}$ | 1.57 | 1.60 | 1.54 | 1.57 | 1.57 | 1.55 | Hz |
| | $\beta_{uy}$ | 36.4 | 24.6 | 45.4 | 36.4 | 36.4 | 24.9 | − |
| $G_{\delta_H\varphi}$ | $V_{uy}^0$ | 0.025 | 0.025 | 0.023 | 0.025 | 0.025 | 0.033 | − |
| | $V_{uy}^{max}$ | 0.049 | 0.039 | 0.058 | 0.049 | 0.049 | 0.043 | − |
| | $\omega_0^{uy}$ | 1.21 | 0.99 | 1.28 | 1.21 | 1.21 | 1.05 | Hz |
| | $\beta_{uy}$ | 2.48 | 2.45 | 2.60 | 2.48 | 2.48 | 2.79 | − |
| $G_{\delta_H\dot\varphi}$ | $V_{uy}^0$ | 0.050 | 0.052 | 0.046 | 0.050 | 0.050 | 0.064 | 1/s |
| | $V_{uy}^{max}$ | 0.33 | 0.25 | 0.40 | 0.33 | 0.33 | 0.27 | 1/s |
| | $\omega_0^{uy}$ | 1.33 | 1.34 | 1.36 | 1.33 | 1.33 | 1.28 | Hz |
| | $\beta_{uy}$ | 6.72 | 4.69 | 8.68 | 6.72 | 6.72 | 4.29 | − |
| $G_{\delta_H\ddot\varphi}$ | $V_{uy}^0$ | 0.090 | 0.102 | 0.086 | 0.090 | 0.090 | 0.111 | 1/$s^2$ |
| | $V_{uy}^{max}$ | 2.37 | 1.78 | 2.88 | 2.37 | 2.37 | 1.91 | 1/$s^2$ |
| | $\omega_0^{uy}$ | 1.41 | 1.46 | 1.42 | 1.41 | 1.41 | 1.39 | Hz |
| | $\beta_{uy}$ | 26.4 | 17.4 | 33.7 | 26.4 | 26.4 | 17.1 | − |
| $G_{a_y\varphi}$ | $V_{uy}^0$ | 0.19 | 0.18 | 0.14 | 0.16 | 0.22 | 0.20 | |
| | $V_{uy}^{max}$ | 0.39 | 0.30 | 0.38 | 0.34 | 0.46 | 0.26 | |
| | $\omega_0^{uy}$ | 1.36 | 0.92 | 1.40 | 1.36 | 1.36 | 1.22 | Hz |
| | $\beta_{uy}$ | 2.24 | 2.26 | 2.70 | 2.24 | 2.24 | 2.57 | − |
| $G_{a_y\dot\varphi}$ | $V_{uy}^0$ | 0.37 | 0.38 | 0.29 | 0.32 | 0.43 | 0.38 | |
| | $V_{uy}^{max}$ | 2.82 | 1.98 | 2.80 | 2.45 | 3.32 | 1.76 | |
| | $\omega_0^{uy}$ | 1.45 | 1.44 | 1.47 | 1.45 | 1.45 | 1.33 | Hz |
| | $\beta_{uy}$ | 7.71 | 5.18 | 9.71 | 7.71 | 7.71 | 4.68 | − |
| $G_{a_y\ddot\varphi}$ | $V_{uy}^0$ | 0.64 | 0.73 | 0.53 | 0.56 | 0.76 | 0.65 | |
| | $V_{uy}^{max}$ | 21.0 | 14.8 | 21.0 | 18.2 | 24.7 | 12.4 | |
| | $\omega_0^{uy}$ | 1.50 | 1.52 | 1.53 | 1.50 | 1.50 | 1.40 | Hz |
| | $\beta_{uy}$ | 32.5 | 20.4 | 40.0 | 32.5 | 32.5 | 19.0 | − |

*Correlation Analysis*

Both auto- and cross-correlations were evaluated for the three data sets obtained through the objective measurements and the subjective assessments (subjective ratings of both liking and intensity, and objective parameters). The resulting auto-correlation coefficients of the quasi-objective ratings are listed in table 5, where a blue background colour equals a correlation coefficient of +1, a white background colour equals a correlation coefficient of 0 and a correlation coefficient of -1 equals a red background colour. The same colour scheme was also used in tables 6 to 8. Low correlation coefficients are an indication of the test subjects perceiving the variation of the assessment criteria as independent which is consistent with the artificial creation of the variants.

Table 5: Auto-correlation coefficients of the quasi-objective ratings.

|  |  | Criteria (quasi-objective) | | | | | |
|---|---|---|---|---|---|---|---|
|  |  | **RAL** | **RAH** | **TDL** | **TDH** | **IRM** | **ROS** |
| **Criteria (quasi-objective)** | **RAL** | 1.00 |  |  |  |  |  |
|  | **RAH** | 0.00 | 1.00 |  |  |  |  |
|  | **TDL** | **0.98** | -0.09 | 1.00 |  |  |  |
|  | **TDH** | -0.01 | 0.27 | 0.12 | 1.00 |  |  |
|  | **IRM** | 0.60 | 0.73 | 0.56 | 0.30 | 1.00 |  |
|  | **ROS** | 0.52 | 0.66 | 0.54 | 0.47 | **0.86** | 1.00 |

Overall only two statistically significant correlation coefficients were found, one between RAL and TDL and the other between IRM and ROS. Because time delays often are harder to perceive and compare than other criteria (which was also stated by several test subjects during the study) and due to the small number of distinguished variants regarding TDL, it is not surprising the test subjects unconsciously based their rating of TDL on RAL with the latter being the only other lower frequency criterion.

The other significant correlation was found between IRM and ROS which could either indicate that the intensity of both initial and final body roll motion was perceived similarly for the assessed variants or that one criterion was rated according to the other. Given the above-mentioned small difference between the variants in respect to ROS and partly for IRM as well, the first assumption seems more probable.

Table 6: Auto-correlation coefficients of the liking ratings.

|  |  | Criteria (subjective) | | | | | | |
|---|---|---|---|---|---|---|---|---|
|  |  | **RAL** | **RAH** | **TDL** | **TDH** | **IRM** | **ROS** | **OR** |
| Criteria (subjective) | **RAL** | 1.00 | | | | | | |
|  | **RAH** | -0.19 | 1.00 | | | | | |
|  | **TDL** | **0.87** | -0.48 | 1.00 | | | | |
|  | **TDH** | -0.01 | **0.95** | -0.24 | 1.00 | | | |
|  | **IRM** | **0.89** | 0.15 | 0.70 | 0.31 | 1.00 | | |
|  | **ROS** | 0.58 | 0.43 | 0.38 | 0.60 | **0.83** | 1.00 | |
|  | **OR** | 0.70 | 0.39 | 0.53 | 0.57 | **0.93** | **0.94** | 1.00 |

The auto-correlation results of the subjective ratings are presented in table 6. There were six significant linear correlations between the liking ratings of the assessment criteria. The large majority of ratings did not correlate, which could be expected from the isolated variation of the roll dynamics aspects. The correlation of RAL and TDL is not surprising given the correlation of the corresponding quasi-objective ratings. Because the quasi-objective ratings of RAH and TDH do not correlate, the correlation of their subjective ratings suggests that the time delay criterion was rated according to RAH. Due to the correlation of the intensity ratings of IRM and ROS, the auto-correlation of the corresponding liking ratings is not surprising. The correlation of OR with both the criteria IRM and ROS could be explained either by one of the individual aspects dominating the subjective impression of the variants or by the individual ratings being more difficult to assess so the test subjects rated them according to the overall impression of the vehicle variant.

Table 7 shows the coefficients of the correlation of the subjective ratings with the quasi-objective ratings. Ideally, there should be significant correlations on the main diagonal, which would indicate consistency between the ratings of liking and intensity for each criterion and confirm the assessment quality of the test subjects. The only criterion for which this is not true is the higher frequency time delay TDH which might again point towards the difficulty of perceiving and comparing small differences between time delays. The correlation of the quasi-objective ratings of IRM and ROS with the subjective ratings of ROS and IRM can be explained from the corresponding auto-correlation results of both the subjective and the quasi-objective ratings shown in tables 5 and 6, respectively.

Table 7: Cross-correlation coefficients of the subjective and the quasi-objective ratings.

|  |  | **Criteria (quasi-objective)** | | | | | |
|---|---|---|---|---|---|---|---|
|  |  | **RAL** | **RAH** | **TDL** | **TDH** | **IRM** | **ROS** |
| **Criteria (subjective)** | **RAL** | **-0.79** |  |  |  |  |  |
|  | **RAH** | 0.11 | **-0.88** |  |  |  |  |
|  | **TDL** | **-0.90** | 0.27 | **-0.90** |  |  |  |
|  | **TDH** | -0.14 | **-0.89** | -0.10 | -0.56 |  |  |
|  | **IRM** | -0.74 | -0.35 | -0.75 | -0.23 | **-0.88** |  |
|  | **ROS** | -0.51 | -0.55 | -0.55 | -0.45 | **-0.83** | **-0.97** |
|  | **OR** | -0.69 | -0.48 | -0.73 | -0.49 | **-0.91** | **-0.92** |

Furthermore, based on the correlation results of the subjective ratings and the objective CVs, individual CVs for each criterion for an approximate estimation of its subjective impression can be selected. For this selection it is important to consider the correlation coefficients of both liking and intensity ratings of each CV because the selection of a CV does not make sense if the correlation between CV and liking rating is strong while the correlation between CV and intensity rating is not and vice versa. Any of these two cases would be a strong indication of a random correlation. Basically, the intensity ratings are used to increase the reliability of the selected CVs.

The best individual correlation pair for each criterion from the study data is shown in table 8. Overall, CVs correlating highly with both liking and intensity ratings could be found for every criterion except for RAL and TDL. Because the perceptible variation of RAL and TDL between the variants was not very high, it does not surprise that the coefficients of the correlation of the intensity ratings and the CVs for RAL and TDL are smaller than for the other criteria.

Table 8: Strongest individual correlations for each criterion.

| Criterion | CV | Correlation coefficients | |
|---|---|---|---|
| | | Liking | Intensity |
| RAL | $V^0_{a_y\ddot{\varphi}}$ | -0.84 | 0.57 |
| RAH | $V^{0.9}_{\delta_H\varphi}$ | -0.96 | 0.91 |
| TDL | $V^0_{\delta_H\ddot{\varphi}}$ | -0.85 | 0.54 |
| TDH | $\phi^{0.9}_{M_H\varphi}$ | -0.79 | 0.84 |
| IRM | $\beta_{M_H\varphi}$ | -0.87 | 0.91 |
| ROS | $\beta_{M_H\varphi}$ | -0.80 | 0.76 |
| OR | $\beta_{M_H\varphi}$ | -0.87 | - |

*Regression Analysis*

Finally, linear regression models were identified using a stepwise regression approach with the objective CVs of the seven variants as predictor variables $x_j$ and the subjective ratings as response variables $y_i$:

$$y_i = f(x_j) \qquad (4)$$

Additional predictor terms were added to the equations only if the value of $R^2_{\text{adj}}$ was increased by at least 0.05 to avoid overfitting while an $R^2_{\text{adj}}$ of more than 0.90 was accepted as sufficient. The results of the regression analysis are presented in equations 5 to 11 with the corresponding value ranges below each equation.

$$RAL = 11.4 - 1.77 \cdot \beta_{\delta_H\varphi} 0.92 \cdot \omega^0_{M_H\varphi} - 0.20 \cdot V^{max}_{M_H\ddot{\varphi}} \qquad (5)$$

$$2.45 \leq \beta_{\delta_H\varphi}/[-] \leq 2.79$$

$$0.92 \leq \omega^0_{M_H\varphi}/[Hz] \leq 1.40$$

$$1.76 \leq V^{max}_{M_H\ddot{\varphi}}/[°/sNm] \leq 3.32$$

$$RAH = 9.5 - 0.82 \cdot V_{a_y\dot{\varphi}}^{max} \tag{6}$$

$$2.35 \leq V_{a_y\dot{\varphi}}^{max}/[°s/m] \leq 3.73$$

$$TDL = 9.3 - 20.69 \cdot V_{\delta_H\ddot{\varphi}}^{0} \tag{7}$$

$$0.086 \leq V_{\delta_H\ddot{\varphi}}^{0}/[1/s^2] \leq 0.11$$

$$TDH = 8.9 - 1.05 \cdot V_{a_y\dot{\varphi}}^{max} + 1.97 \cdot \omega_{\delta_H\varphi}^{0} - 1.52 \cdot V_{M_H\ddot{\varphi}}^{0} \tag{8}$$

$$2.35 \leq V_{a_y\dot{\varphi}}^{max}/[°s/m] \leq 3.73$$

$$0.99 \leq \omega_{\delta_H\varphi}^{0}/[Hz] \leq 01.28$$

$$0.53 \leq V_{M_H\ddot{\varphi}}^{0}/[°/s^2Nm] \leq 0.76$$

$$IRM = 11.6 + 3.14 \cdot V_{a_y\varphi}^{1.2} - 2.68 \cdot V_{M_H\varphi}^{0.6} - 2.00 \cdot \beta_{M_H\varphi} \tag{9}$$

$$0.32 \leq V_{a_y\varphi}^{1.2}/[°s^2/m] \leq 0.48$$

$$0.22 \leq V_{M_H\varphi}^{0.6}/[°/Nm] \leq 0.32$$

$$2.24 \leq \beta_{M_H\varphi}/[-] \leq 2.70$$

$$ROS = 2.8 - 0.019 \cdot \phi_{\delta_H\varphi}^{1.5} - 2.94 \cdot V_{M_H\varphi}^{0.6} - 1.27 \cdot \beta_{M_H\varphi} \tag{10}$$

$$-99 \leq \phi_{\delta_H\varphi}^{1.5}/[°] \leq -86$$

$$0.22 \leq V_{M_H\varphi}^{0.6}/[°/Nm] \leq 0.32$$

$$2.24 \leq \beta_{M_H\varphi}/[-] \leq 2.70$$

$$OR = 3.6 - 0.021 \cdot \phi_{a_y\varphi}^{1.5} - 1.43 \cdot \beta_{M_H\varphi} - 1.18 \cdot V_{M_H\ddot{\varphi}}^{0} \tag{11}$$

$$-15 \leq \phi_{a_y\varphi}^{1.5}/[°] \leq -3$$

$$2.24 \leq \beta_{M_H\varphi}/[-] \leq 2.70$$

$$0.53 \leq V_{M_H\ddot{\varphi}}^{0}/[°/s^2Nm] \leq 0.76$$

The values of $R_{adj}^2$ of equations 5 to 11 are 1.00, 0.93, 0.65, 0.97, 1.00, 0.79 and 0.99,

respectively. The objective measurement data was then inserted into these equations for back-testing purposes, i.e. to check the consistency of the predicted ratings and the original ratings from the study. The resulting rating predictions are depicted in figure 5.

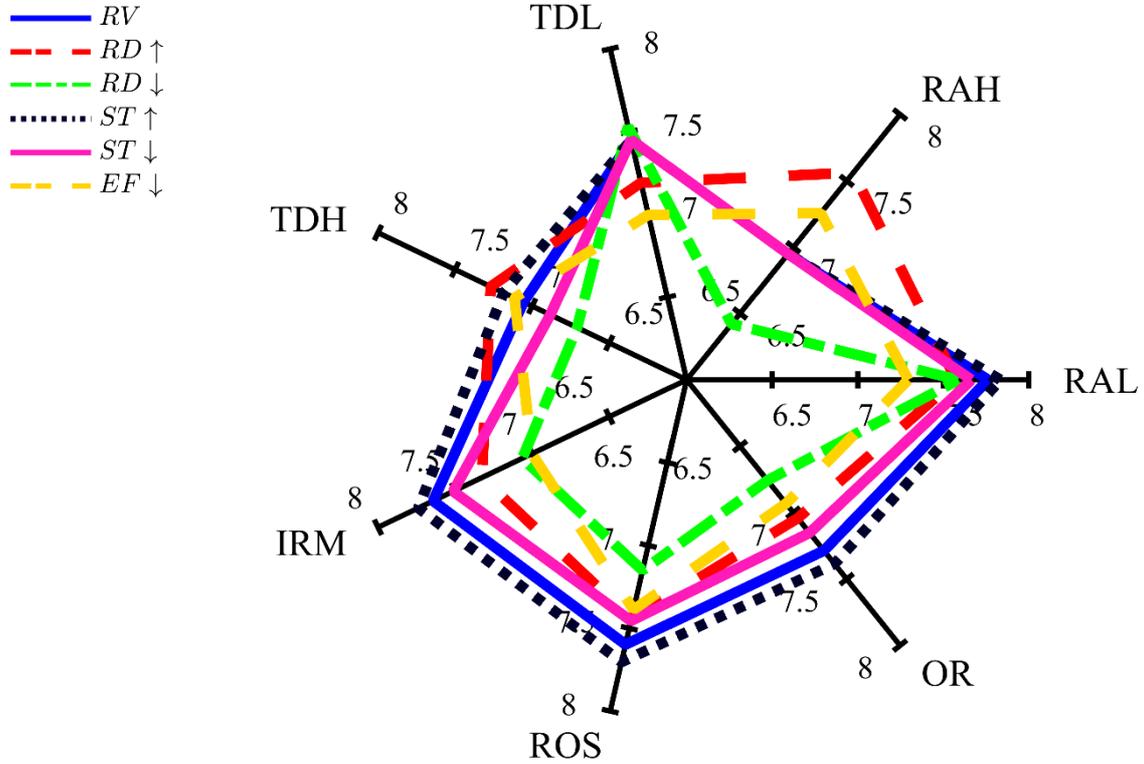

Figure 5: Predictions for the BMW X5 variants.

To measure the prediction error for each criterion $i$, the root mean squared error (RMSE) defined by

$$RMSE \coloneqq \sqrt{\sum_i (y_{i,orig} - y_{i,pred})^2}$$

is used, where $y_{orig}$ is the original rating and $y_{pred}$ is the predicted rating of each criterion $i$. The highest RMSE is 0.3 for TDL followed by 0.2 for RAH, and 0.1 for ROS. The RMSEs for the remaining criteria are all below 0.1. Thus, a very good fit of the original and the predicted ratings could be achieved for the given vehicle data and their corresponding subjective ratings.

*Validation of the Prediction Models*

The results of the correlation and regression analysis were then additionally validated using previously recorded measurement data of several other vehicles. Their frequency domain data is shown in figure 6. The dataset of vehicles comprised both extremes of vehicles available to market, ranging from a super sports car to a super luxury sedan, with a sports SUV, a sedan and a coupe in between them. The predicted subjective ratings of the five variants are shown in figure 7.

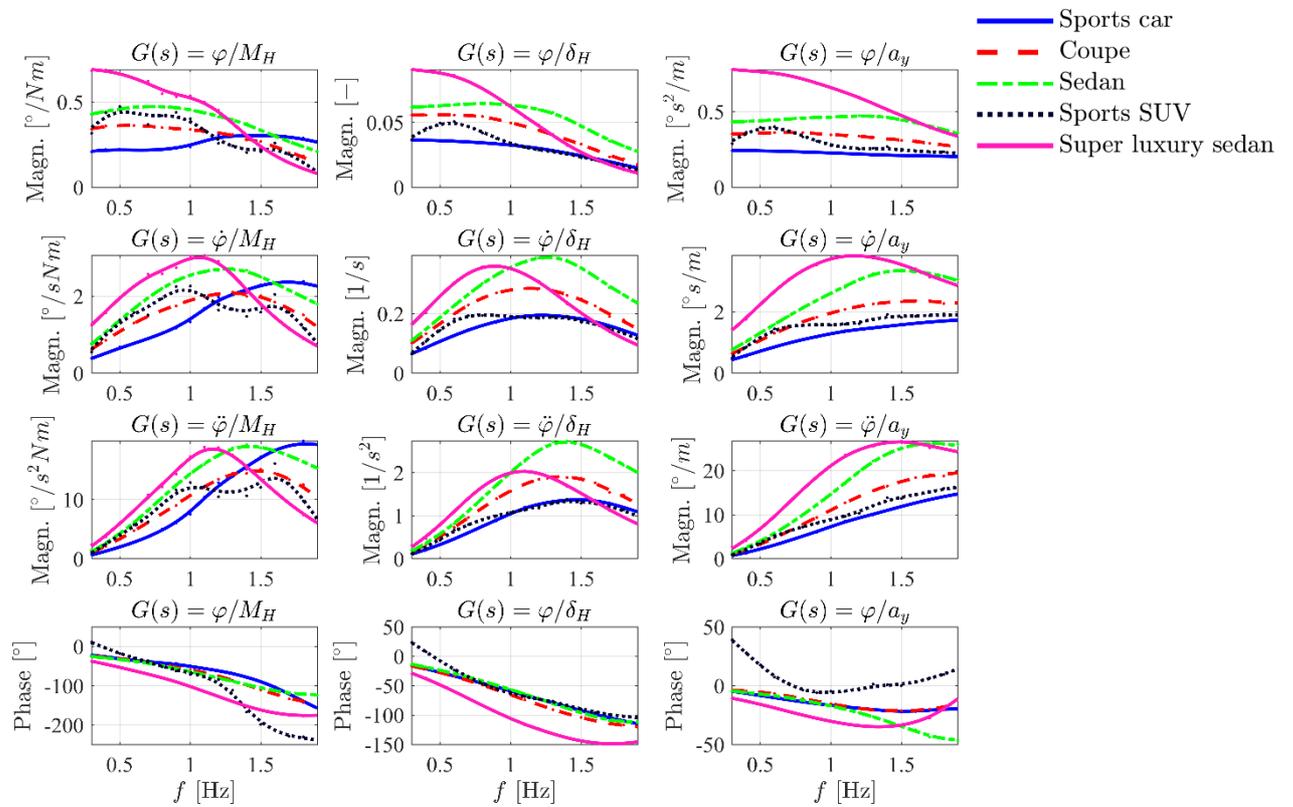

Figure 6: Frequency-domain data used for model validation.

The relative order of the predicted ratings for each criterion seems plausible for the most part. The super sports car with barely any roll motion received the highest ratings whereas the super luxury sedan received the lowest ratings. The coupe and the sedan received pretty similar ratings, with the ratings of most of their criteria alternating between the second- and third-best ratings. The sports SUV was predicted to be rated fourth-best overall, even though its criteria RAH, TDL and TDH were calculated to be on par with the super sports car's ones. Because some of the objective CVs of the validation vehicles lie outside the range of the objective data of the variants from the original study, the predictions from the regression models do not necessarily end up within the original scale from 1 to 10 which is the case here. Nevertheless, the relative order of the predictions should still hold true.

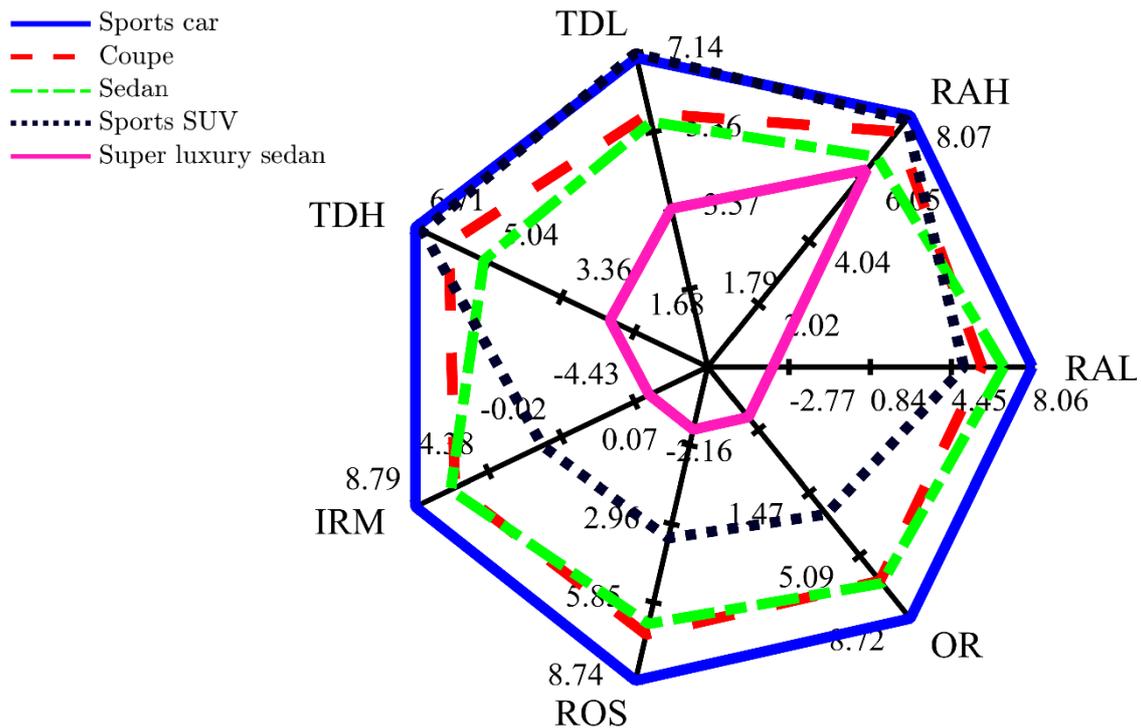

Figure 7: Predictions for the validation data set.

**Discussion**

The correlation analysis provided a set of parameters with relatively high correlation coefficients, even though the correlations of the quasi-objective ratings of RAL and TDL could be higher. Most of the CVs match the intuitive understanding of how the individual aspect could be predicted, except for ROS. The low coefficients of the quasi-objective correlation of RAL and TDL can be attributed to the small difference of the variants regarding these criteria and should be investigated further.

Based on the available frequency-domain parameters, the regression models almost perfectly predicted the subjective ratings from the study even though up to three predictor variables are required. The influence of the steering torque is taken into account as well with five of the seven models using at least one steering torque CV. This indicates that, when analysing the perception of vehicle roll dynamics, it is important to consider the complete chain of effects and not just the relationship between lateral acceleration and roll angle. The rough validation performed at the end of this paper shows that trends are captured by the models for the most part but there are still some effects that were not covered in this study, mainly because the range of vehicle dynamics covered by the original variants could be larger.

Next, the prediction models need to be expanded to increase their robustness and the reliability of the predicted ratings. Ideally, another study with several vehicles spanning a broad range of potential vehicle behaviours is conducted so that the results of both studies can be combined to increase the validity of the models. Additionally, time-domain CVs could be used, which might further improve the prediction quality of the rating criteria.